\documentclass{sig-alternate}
\usepackage{amsmath}
\usepackage{algorithm}
\usepackage{algorithmic}
\usepackage{multirow}
\usepackage{graphicx}
\usepackage{epsfig}

\begin{document}

\title{Identifying Relevant Document Facets for Keyword-Based Search Queries}

\numberofauthors{1}
\author{
\alignauthor
Lanbo Zhang
\email{lanbozhang@gmail.com}
}

\maketitle

\begin{abstract}

As structured documents with rich metadata (such as products, movies, etc.) become increasingly prevalent, searching those documents has become an important IR problem. Although advanced search interfaces are widely available, most users still prefer to use keyword-based queries to search those documents. Query keywords often imply some hidden restrictions on the desired documents, which can be represented as document facet-value pairs. To achieve high retrieval performance, it's important to be able to identify the relevant facet-value pairs hidden in a query. In this paper, we study the problem of identifying document facet-value pairs that are relevant to a keyword-based search query. We propose a machine learning approach and a set of useful features, and evaluate our approach using a movie data set from INEX.

\end{abstract}




\section{Introduction}

\begin{figure}
\centering
\includegraphics[scale=0.5]{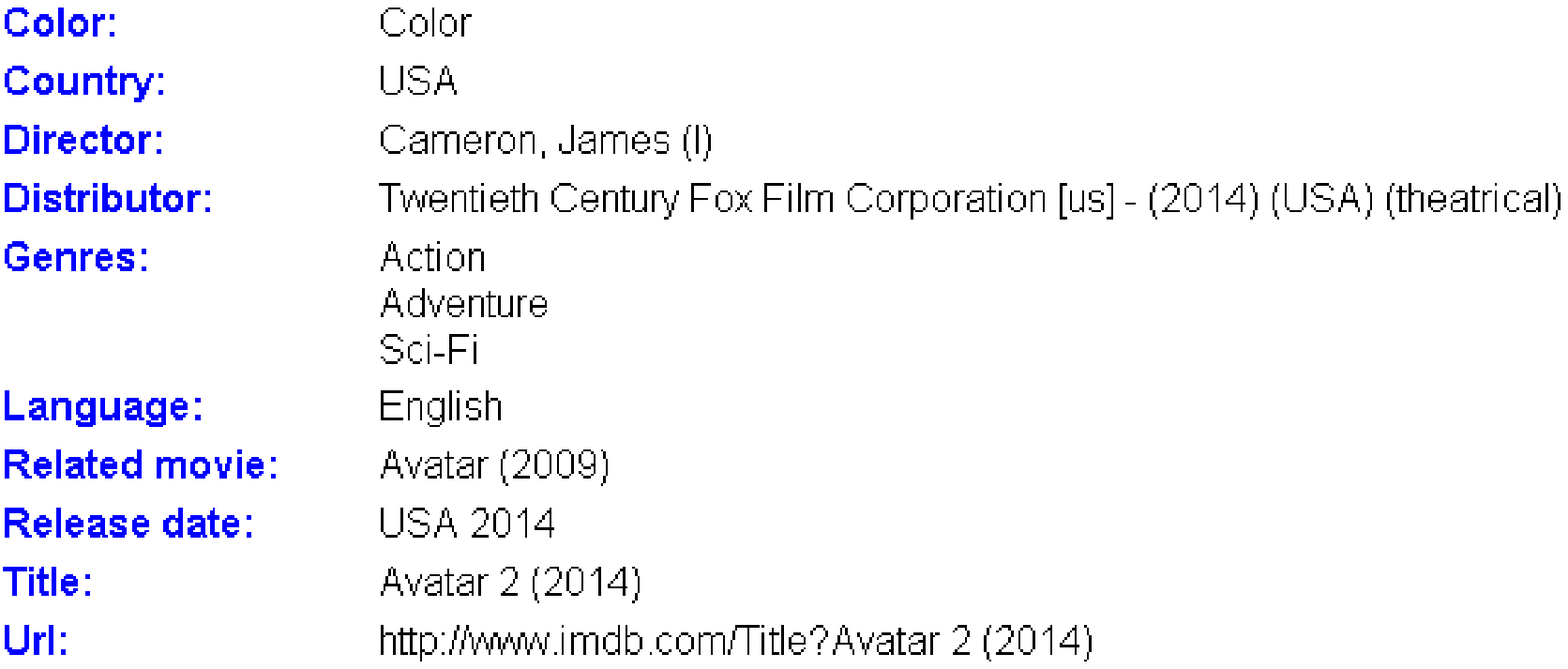}
\caption{An example of structured document that describes a movie}\label{fig:movie}
\end{figure}

Structured documents with rich metadata are increasingly prevalent on the Internet. A large portion of structured documents are those representing various types of entities, such as products, movies, images, businesses, jobs, people, etc. In these documents, each metadata field characterizes a specific facet of the entity, and may be assigned with one or several values. In this paper, we call each metadata field a document \textbf{facet}, a metadata field assigned with a particular value a \textbf{facet-value pair} (FVP). Figure \ref{fig:movie} shows a movie document, where the bold words are facets, each of which is followed by the value(s) of the facet. In this document, the facet ``genre'' has three values: ``Action'', ``Adventure'', and ``Sci-Fi'', which correspond to three facet-value pairs respectively: ``genre: Action'', ``genre: Adventure'', and ``genre: Sci-Fi''.

A keyword-based query searching for structured documents usually implies a set of facet-value pairs that jointly define the information need behind the query. For example, query ``15-inch silver laptop by Lenovo'' implies the relevant products must meet the following restrictions: ``product category: laptop'', ``maker: Lenovo'', ``color: silver'', ``screen size: 15-inch''. If the system can identify the relevant facet-value pairs behind a query, the system will be able to return more accurate documents to the user.

In this paper, we study the problem of identifying relevant facet-value pairs from query keywords. The central question is: given a keyword query, how to find the relevant facet-value pairs that the retrieved documents should have? To answer this question, we propose a learning-based approach and a set of features, and evaluate our approach on a movie data set from INEX. Experiment results show that the learning-based approach can effectively find relevant facet-value pairs and the proposed features are very useful.

In our previous work \cite{Zhang:2010}, we proposed several heuristic approaches for ranking facet-value pairs based on relevance to the query. In this paper, our approach is different in two aspects. First, we focus on highly structured documents where metadata dominates each document. Secondly, we use a learning-based approach for FVP ranking. Compared with heuristic approaches, learning-based approaches usually perform better because of being able to combine multiple relevance signals. In fact, some of the features we propose in this paper are equivalent to the best approaches proposed in \cite{Zhang:2010}, and our experiment results show that the learning-based approach can dramatically improve FVP ranking performance (Table \ref{tab:fvp-perf}).

\section{A Learning-Based Approach}

To obtain labeled data for model training, we hire a human assessor to provide relevant judgments between queries and facet-value pairs. As for the learner, we use the Gradient Boosted Trees (GBT) \cite{Friedman2001}, which is able to handle deep interactions among features, and has been shown to perform well in many IR tasks such as learning to rank \cite{Li2006}.

\subsection{Features}

We propose a number of features that measure the relevance degree between query and facet-value pair. These features can be categorized into seven categories based on what type of information is used. Table \ref{tab:features} summarizes all the features.

\subsubsection{Query Features}

This type of features only depends on the query. We use two features: the query length (number of words), and the average IDF of all query words. The average IDF is used to measure the uniqueness of a query.

\subsubsection{Facet Features}

This type of features only depend on the facet. The first (F.Type) is a categorical feature that identifies the facet (the number of unique facets is usually small). The second feature (F.NumValues) is the number of unique values the facet has in the whole corpus. The third feature (F.NumOccrs) is the total number of occurrences of all facet-value pairs of this facet in the whole corpus.

\subsubsection{Value Features}

This type of features only depend on the value. Two features are used: V.Length is the number of words contained in the value, and V.AvgIDF is the average IDF of all value words.

\subsubsection{FVP Features}

This type of features depend on the facet-value pair. P.NumDocs is the number of documents with this facet-value pair. P.IDF is the Inverse Document Frequency of this FVP.

\subsubsection{Query-Facet Features}

This type of features measure the similarity between the query and the facet. QF.TFIDF is the TFIDF score between the query and the facet. This feature might be useful based on the intuition that users might use the facet name to express the faceted constraint on returned documents. For example, the query ``movies directed by James Cameron'' is related to the facet ``director''.

\subsubsection{Query-Value Features}

This type of features measure the similarity between the query and the value. We use four features based on four traditional IR scoring methods. QV.BM25 and QV.TFIDF are the BM25 and TFIDF scores between the query and the value. QV.SIDF is different from QV.TFIDF, and ignores the term frequency part. QV.CosSim is the cosine similarity, where the query and value vectors are calculated using the TFIDF weighting method. Comparing these four features, QV.BM25 and QV.CosSim have penalty for long values while the other two do not.

\subsubsection{Query-FVP Features}

This type of features depend on both the query and the facet-value pair, which are mainly based on the frequency of the FVP occurring in the top retrieved documents. QP.DFN measures how many documents in the top N retrieved documents have the FVP. We set N = 10, 100, 1000, and the number of all retrieved documents respectively. QP.DFIDFN is the product of QP.DFN and the IDF of the FVP. This group of features might be useful based on the intuition that a facet-value pair occurring frequently in the top retrieved documents while less frequently in the whole corpus are more likely to be relevant to the query.

\begin{table}[ht]
\caption{Features for ranking facet-value pairs}
\label{tab:features}
\begin{center}
\begin{tabular}{|p{1cm}|p{2cm}|p{4.5cm}|}
\hline
\textbf{Type} & \textbf{ID} & \textbf{Detail} \\
\hline
\multirow{2}{1cm}{Query} & Q.Length & Number of words in the query \\
\cline{2-3}
& Q.AvgIDF & Average IDF of query words \\
\hline
\multirow{3}{1cm}{Facet} & F.Type & The facet type (categorical) \\
\cline{2-3}
& F.NumValues & Number of unique values of this facet \\
\cline{2-3}
& F.NumOccrs & Number of occurrences of all values of this facet \\
\hline
\multirow{2}{1cm}{Value} & V.Length & Number of words in the value \\
\cline{2-3}
& V.AvgIDF & Average IDF of value words \\
\hline
\multirow{2}{1cm}{FVP} & P.NumDocs & Number of documents containing this FVP \\
\cline{2-3}
& P.IDF & IDF of this FVP \\
\hline
\multirow{1}{1cm}{Query-Facet} & QF.TFIDF & TFIDF score between the query and the facet name \\
\hline
\multirow{4}{1cm}{Query-Value} & QV.TFIDF & TFIDF score between the query and the value \\
\cline{2-3}
& QV.BM25 & BM25 score between the query and the value \\
\cline{2-3}
& QV.SIDF & Sum of IDFs of the overlapped words between the query and the value \\
\cline{2-3}
& QV.CosSim & Cosine similarity between the query and the value \\
\hline
\multirow{6}{1cm}{Query-FVP} & QP.DF10 & FVP frequency in the top 10 retrieved documents \\
\cline{2-3}
& QP.DFIDF10 & QP.DF10 * IDF of the FVP \\
\cline{2-3}
& QP.DF100 & FVP frequency in the top 100 retrieved documents \\
\cline{2-3}
& QP.DFIDF100 & QP.DF100 * IDF of the FVP \\
\cline{2-3}
& QP.DF1000 & FVP frequency in the top 1000 retrieved documents \\
\cline{2-3}
& QP.DFIDF1000 & QP.DF1000 * IDF of the FVP \\
\cline{2-3}
& QP.DFAll & FVP frequency in all retrieved documents \\
\cline{2-3}
& QP.DFIDFAll & QP.DFAll * IDF of the FVP \\
\hline
\end{tabular}
\end{center}
\end{table}

\section{Experiments}

\subsection{Data Set}

We use the data set from the data-centric track of INEX 2010 \cite{Geva10inex2010}. This data set consists of: 1) the IMDB movie collection with 1,594,513 movies; 2) 26 query topics (title, description, and narrative) created by the track participants.

To prepare labeled facet-value pairs for model training, we hire a human assessor to provide relevance judgments on query-FVP pairs. To obtain FVP candidates, all FVPs are ranked based on the BM25 score between the query and the value (feature QV.BM25 in Table \ref{tab:features}), and the top 100 FVPs of each query are kept for relevance judging. As a result, we have a total number of 2600 query-FVP relevance judgments, among which there are 148 relevant ones. A query example and the relevant facet-value pairs labeled by the human assessor are shown in Table \ref{tab:query}.

\begin{table}
\caption{A query example in the data set}
\label{tab:query}
\begin{center}
\begin{tabular}{|p{1.5cm}|p{6cm}|}
\hline
Title: & Comedy Woody Allen Scarlett Johansson \\
\hline
Description: & Comedy movies directed by Woody Allen and acted by Scarlett Johansson. \\
\hline
Narrative: & I am looking for the comedy movies directed by Woody Allen and acted by Scarlett Johansson. \\
\hline
\multirow{3}{1.5cm}{Relevant FVPs:} & Genre: Comedy \\
& Director: Woody Allen \\
& Actor: Scarlett Johansson \\
\hline
\end{tabular}
\end{center}
\end{table}

\subsection{Experiment Settings}

We use the package from \cite{GBM2006} for training gradient boosted trees. To determine the best number of trees in GBT, we use the best MAP instead of the smallest error on the validation set. Regarding the parameters of GBT, we use the Bernoulli distribution, a maximum of 3000 trees, an interaction depth of 5, a minimum number of observations of 10 in each tree node, and a shrinkage parameter of 0.01. We use 10-fold cross validation to evaluate our approach, and the average performance on all folds will be reported.

We use the BM25 algorithm implemented in Lemur as the document retrieval approach throughout our experiments. Before scoring a document, we remove all XML tags, and treat it as an unstructured document.

In our approaches, we assume only the title part of each query is available, since keywords in the title part are typical queries in commercial search engines. The query descriptions and narratives are only used to help the human assessor make relevance judgments of queries and facet-value pairs.

\section{Experiment Results}

Table \ref{tab:fvp-perf} shows the ranking performance of each FVP-ranking approach. GBT is the learning-based approach that uses all features in Table \ref{tab:features}. All the other 12 approaches are based on individual features. We didn't use the other features because they do not measure the relevance degree between query and FVP directly, and thus are not suitable for FVP ranking individually.

\begin{table}[ht]
\caption{Performances of different FVP-ranking approaches. GBT is the learning-based approach, and the other approaches use individual features. GBT significantly outperforms all the other approaches under a paired t-test (p-value < 0.05).}
\label{tab:fvp-perf}
\begin{center}
\begin{tabular}{|l|l|l|l|l|}
\hline
\bf{Approach} & \bf{MAP} & \bf{R-Prec} & \bf{P@5} & \bf{P@R=1} \\
\hline
\bf{QV.TFIDF} & 0.18 & 0.09 & 0.08 & 0.11 \\
\bf{QV.SIDF} & 0.18 & 0.15 & 0.14 & 0.10 \\
\bf{QP.DF10} & 0.30 & 0.25 & 0.22 & 0.23 \\
\bf{QP.DFIDF10} & 0.32 & 0.28 & 0.22 & 0.24 \\
\bf{QP.DFAll} & 0.33 & 0.21 & 0.30 & 0.21 \\
\bf{QP.DFIDFAll} & 0.33 & 0.21 & 0.30 & 0.21 \\
\bf{QV.CosSim} & 0.37 & 0.24 & 0.33 & 0.23 \\
\bf{QV.BM25} & 0.42 & 0.29 & 0.34 & 0.28 \\
\bf{QP.DF1000} & 0.48 & 0.38 & 0.29 & 0.36 \\
\bf{QP.DFIDF1000} & 0.51 & 0.43 & 0.29 & 0.38 \\
\bf{QP.DFIDF100} & 0.53 & 0.43 & 0.29 & 0.37 \\
\bf{QP.DF100} & 0.53 & 0.44 & 0.29 & 0.37 \\
\bf{GBT} & \textbf{0.73} & \textbf{0.61} & \textbf{0.45} & \textbf{0.55} \\
\hline
\end{tabular}
\end{center}
\end{table}

\begin{figure}
\centering
\epsfig{file=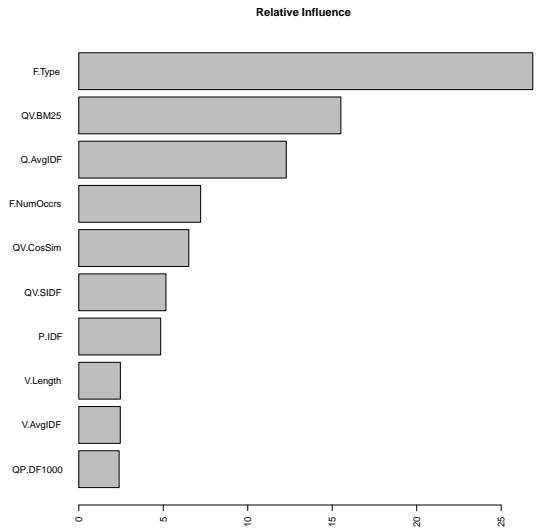, scale=0.3}
\caption{Top 10 features with highest relative influence in GBT}\label{fig:influence}
\end{figure}

Based on Table \ref{tab:fvp-perf}, we have the following findings:

\itemize {
\item GBT dramatically outperforms all individual features. This demonstrates the superiority of the learning-based approach, and indicates the features we proposed are complementary with each other.

\item QP (Query-FVP) features generally perform better than QV (Query-Value) features, which means the frequency among top retrieved documents (measured by QP features) is a stronger signal than the pure text match between the value and the query.

\item Among all QP features, QP.DF100 and QP.DFIDF100 perform the best, which means 100 might be a reasonable cutoff on top retrieved documents on the data set we use.

\item Among all QV features, QV.BM25 and QV.CosSim outperform QV.TFIDF and QV.IDF significantly. Note that the major difference between these features is that QV.BM25 and QV.CosSim normalize term frequency based on the value length of the FVP while the other two do not. Given the dramatically different performances, we can conclude that length normalization is very important for FVP ranking, and this might be generalized to other short-text-ranking problems as well.
}

The top 10 features with highest relative influences output by the GBM package are shown in Figure \ref{fig:influence}.

\section{Conclusions}

We study the problem of identifying relevant facet-value pairs for keyword-based search queries. This task is important because it can be applied in many applications, including structured document retrieval/filtering \cite{Zhang:2011:FSD:2009916.2010003,Zhang:2013:CFS:2604906}, interactive retrieval \cite{Zhang:2010,zhang2009ucsc}, structured document summarization (snippet generation) \cite{Zhang:2012:SHS:2348283.2348306}, etc. We proposed a supervised learning approach for this problem. The proposed approach is evaluated on a movie data set. Experiment results show the proposed approach can identify relevant facet-value pairs effectively.

\bibliographystyle{abbrv}
\bibliography{sigir12}

\end{document}